\documentclass[twocolumn,showpacs,preprintnumbers,amsmath,amssymb,superscriptaddress,aps]{revtex4}

\usepackage{graphicx}% Include figure files
\usepackage{dcolumn}% Align table columns on decimal point
\usepackage{bm}% bold math

\begin{document}

%\preprint{cf}

\title{Evidence of Landau Levels and Interactions in Low-lying Excitations of Composite Fermions at $1/3\leq \nu \leq 2/5$}% Force line breaks with \\

\author{Irene Dujovne}
%\email[id61@columbia.edu]{Your e-mail address}
\affiliation{Department of Applied Physics and Applied Math,
Columbia University, New York, NY  10027} \affiliation{Bell Labs,
Lucent Technologies, Murray Hill, NJ 07974}

\author{A. Pinczuk}
\affiliation{Department of Applied Physics and Applied Math,
Columbia University, New York, NY  10027} \affiliation{Bell Labs,
Lucent Technologies, Murray Hill, NJ 07974}
\affiliation{Department of Physics, Columbia University, New York,
NY  10027}
\author{Moonsoo Kang \footnote{Present address: Physics Department, Washington State University, 1245 Webster
Pullman, WA 99164-2814.}} \affiliation{Department of Applied Physics
and Applied Math, Columbia University, New York, NY  10027}
\affiliation{Bell Labs, Lucent Technologies, Murray Hill, NJ 07974}

\author{B.S. Dennis}
\affiliation{Bell Labs, Lucent Technologies, Murray Hill, NJ
07974}

\author{L.N. Pfeiffer}
\affiliation{Bell Labs, Lucent Technologies, Murray Hill, NJ
07974}

\author{K.W. West}
\affiliation{Bell Labs, Lucent Technologies, Murray Hill, NJ
07974}

\date{\today}

\begin{abstract}
Excitation modes in the range $2/5 \geq \nu \geq 1/3$ of the
fractional quantum Hall regime are observed by resonant inelastic
light scattering. Spectra of spin reversed excitations suggest a
structure of lowest spin-split Landau levels of composite fermions
that is similar to that of electrons. Spin-flip energies
determined from spectra reveal significant composite fermion
interactions. The filling factor dependence of mode energies
display an abrupt change in the middle of the range when there is
partial population of a composite fermion level.
\end{abstract}

\maketitle

The states of the fractional quantum Hall effect (FQHE) display
remarkable behaviors that arise from fundamental electron
interactions in two dimensions. Composite fermion (CF)
quasiparticles explain the states with filling factors $\nu =
p/(2np\pm 1)$, where $p$ is the CF filling factor
\cite{jain,heinonen}. In CF's electrons stay apart by binding $2n$
($n=1,2,...$) vortices of the many-body wavefunction. The
odd-denominator FQHE states have integer $p$.  Chern-Simons gauge
fields account for electron interactions so that CF's experience
effective magnetic fields $B^*= B - B_{1/2n} = \pm B/(2np\pm 1)$,
where $B$ is the perpendicular component of the external
field\cite{hlr,kalm}. Composite fermions have spin-split Landau
levels characteristic of charged fermions with spin 1/2, as shown
schematically in the insets to Fig.\ref{fig2/5}(b) and
\ref{fig1/3}(a) for $\nu \lesssim 2/5$ and $\nu \gtrsim 1/3$. In
this simplest of pictures, CF Landau levels resemble those of
electrons. The spacing of lowest same spin levels is represented
as a cyclotron frequency
\cite{hlr,fradkin,du93,simon,park,murthy99,mandal,aoki01}
\begin{equation}
  \omega_c=\frac{e B^*}{c m_{CF}}
  \label{wc}
\end{equation}
where $m_{CF}$ is a CF effective mass. $\omega_c$ is understood as
the energy of the large wavevector ($q\rightarrow\infty$)
inter-Landau level excitation of CF's that represents a
quasiparticle-quasihole pair at large separation \cite{hlr}. In
the main sequence of the FQHE there are $p$ fully occupied CF
Landau levels. Equation \ref{wc} is employed to extract values of
$m_{CF}$ from the field dependence of activated magnetotransport
\cite{du93} and from optically detected microwave absorption
\cite{kukushkinnat}. The existence of a ladder of spin-split
Landau levels of CF's has been deduced from the angular dependence
of magnetotransport at $ 5/3 \geqslant \nu \geqslant 4/3$
\cite{du95} .
\par
Spin-reversed quasiparticle-quasihole pairs have spin-flip
energies that are strongly affected by residual CF interactions
\cite{tapash,longo,aoki,mandalactivation}. States such as 2/5,
with two CF Landau levels fully populated, are spin-polarized
because the spin-flip energies at relatively large fields are
larger than the CF cyclotron frequency. Residual interactions
among composite fermions are of great current interest. While
couplings between CF's are understood as relatively weak, residual
interactions are invoked to interpret fqhe states at  filling
factors in the range 2/5 $>\nu>$ 1/3 when there is partial
population of CF Landau level\cite{pan,izab,scar02}.
\par
Resonant inelastic light scattering methods access the low-lying
excitations of electron liquids of the FQHE
\cite{pinc93,davies97,moonsoo00}. Recent light scattering
experiments at filling factors 1/3 and 2/5 have determined the
energies of rotons at $q\sim 1/l_o$, where $l_o = (\hbar
c/eB)^{1/2}$ is the magnetic length, and of large wavevector
($q\rightarrow\infty$) excitations of CF's \cite{moonsoo01}. The
observed splittings between rotons and $q\rightarrow\infty$ modes
are due to exciton-like bindings in neutral
quasiparticle-quasihole pairs \cite{kallin,haldane,girvin}. These
results, quantitatively explained within CF theory
\cite{moonsoo01,scar00}, suggest that the structure of CF Landau
levels and residual interactions could be explored by light
scattering measurements of collective excitations.
\par
We report resonant inelastic light scattering measurements of
low-lying excitations of the electron liquids in the full range of
filling factors $2/5 \geq \nu \geq 1/3$, where the main FQHE
states are linked to CF's that bind two vortices ($n=1$).
Low-lying excitations are observed at all the magnetic fields
within the range. The experiments enable the study of
quasiparticles in states with partial population of one CF Landau
level ($2 \geq p \geq 1$). They probe the structure of spin-split
CF levels and impact of residual interactions among
quasiparticles.
\par
At filling factors close to 2/5 we observe modes due to spin-flip
(SF) transitions $1\uparrow \rightarrow 0\downarrow$ in which
there are simultaneous changes in spin and CF Landau level quantum
numbers, as shown in the inset to Fig. \ref{fig2/5}(b). Such SF
inter-Landau level collective modes of composite fermions were
recently evaluated for the state at $\nu =$ 2/5\cite{mandal}.
Spin-flip excitations observed at filling factors close to 1/3
when the $1 \uparrow $ level begins to populate confirm the scheme
of spin-split Landau Levels in the inset to Fig. \ref{fig1/3} (a)
and yield a spin-flip energy that reveals significant interactions
between quasiparticles.
\par
The observations of SF modes offer direct insights on the CF
Landau level structure and residual CF interactions. At filling
factors 1/3 and 2/5  the measured spectra are consistent with the
electron-like Landau level structure shown in the insets to
Fig.\ref{fig2/5}(b) and \ref{fig1/3}(a). For filling factors
$\nu\lesssim 2/5$ the SF mode energy has a linear dependence on
effective magnetic field that is consistent with Eq. \ref{wc}. For
larger fields the mode energy becomes independent of field. A
remarkable change in the spectra of low-lying excitations occurs
in the middle of the magnetic field range between 1/3 and 2/5. The
change in character is relatively abrupt, occurring over a narrow
magnetic field interval of $\Delta \nu \lesssim 0.03$. Such
changes in the middle of the filling factor range are regarded as
manifestations of residual quasiparticle interactions in partially
populated CF levels.
\par
The high quality 2D electron system is in a single GaAs quantum
well (SQW) of width $d=330$~\AA\/. The electron density is $\rho =
5.6\times 10^{10}$~cm$^{-2}$. The electron mobility is very high,
reaching a value $\mu\gtrsim7\times10^{6}cm^{2}/Vsec$ at T $ \cong
300mK$. The sample was mounted on the cold finger of a
${}^{3}\text{He}/{}^{4}\text{He}$ dilution refrigerator that is
inserted in the cold bore of a superconducting magnet with windows
for optical access. Cold finger temperatures are variable and as
low as 50~mK. The resonant inelastic light spectra were excited
with a diode laser with photon energies $\omega_{L}$ close to the
fundamental optical gap of the GaAs SQW. The power density was
kept below $10^{-4}$~W/cm$^2$. The spectra were acquired  with a
double Czerny-Turner spectrometer operating in additive mode and a
CCD camera with 15 $\mu$m pixels. The combined resolution with a
30 $\mu$m slit is 0.016 meV. We measured depolarized spectra with
orthogonal incident and scattered light polarizations and
polarized spectra in which polarizations are parallel. Excitation
modes with changes in the spin degree of freedom tend to be
stronger in depolarized spectra, while modes in the charge degree
of freedom tend to be stronger in polarized spectra. A
backscattering geometry shown in the inset to Fig. 1(b) was used
with $\theta \sim 30^o$. The perpendicular component of magnetic
field is $B=B_{T}\cos\theta$, where $B_{T}$ is the total field .
The light scattering wave vector is $k=(2 \omega_{L}/c) \sin\theta
\approx 10^5$~cm$^{-1}$ and $k l_{o} \lesssim 0.1$.
\par
\begin{figure}
\includegraphics{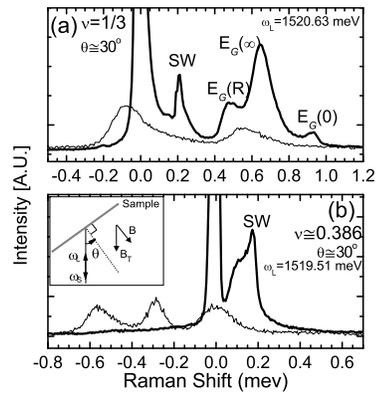}
\caption{\label{lumin} Depolarized inelastic light scattering
spectra (thick lines) and luminescence spectra (thin lines). The
intensity of the luminescence has been re-scaled to adjust a
background intensity. (a) $\nu=1/3$.  (b) $\nu \sim 0.386$. The
inset shows the backscattering geometry, where $\omega_S$ is the
scattered photon energy.}
\end{figure}

In Fig. \ref{lumin} we show the interplay between luminescence and
inelastic light scattering. Resonant light scattering spectra with
photon energies at the fundamental gap are compared with
luminescence spectra excited with higher photon energy. At $\nu$ =
1/3 the light scattering spectrum in Fig. \ref{lumin}(a) displays
four peaks. SW is the long wavelength spin-wave at the Zeeman
energy E$_Z$. The other peaks at E$_G$(R), E$_G$($\infty$) and
E$_G$(0) are charge-density modes (CM) at the roton, at $q
\rightarrow \infty$ and $q \rightarrow 0$ respectively
\cite{moonsoo01}. Figure \ref{lumin} (b) shows spectra taken at
$\nu=0.386$ where the SW mode is dominant. These results reveal
that luminescence intensities have minor impact in resonant light
scattering spectra when photon energies overlap the fundamental
gap.
\par
Figure \ref{fig2/5}(a)  shows light scattering spectra of
low-lying excitations ($\omega \lesssim 0.3meV$) measured at $\nu
\lesssim 2/5$. The spectrum at $\nu=2/5$  shows an asymmetric peak
(SW) at E$_Z=g\mu_B B_T$, where $\mu_B$ is the Bohr magneton and
$g\cong 0.44$ is the Lande factor. SW is the mode in which there
are only changes in the spin quantum number. Spectra at slightly
higher fields, showing better resolved structures, reveal that the
asymmetry of the SW peak at 2/5 is due to a different excitation
mode that is labelled SF$^-$. This mode being below E$_Z$ can only
involve a change in spin because the charge-density modes occur at
energies above E$_Z$\cite{mandal}.  The inset to Fig.
\ref{fig2/5}(a) shows that SW and SF$^-$ modes have the same
polarization selection rules. This also indicates that the mode at
energy below the SW involves a change in the spin degree of
freedom. Figure \ref{fig2/5}(b) shows a calculated dispersion of
SF excitations at $\nu = 2/5$ \cite{mandal}. The calculation
displays a deep roton minimum at energy close to E$_Z$. This
comparison with theory suggests that the mode below E$_Z$ is the
roton of the SF mode. Observations of roton minima are explained
by breakdown of wave vector conservation\cite{davies97,moonsoo01}.
\par
\begin{figure}
\includegraphics{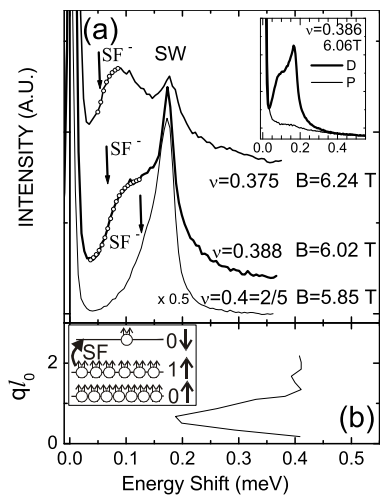}
\caption{\label{fig2/5} a) Inelastic light scattering spectra at
filling factor $\nu \lesssim 2/5$. The dots superimposed on the
measured spectra are fits based on a broadened step density of
states at the roton minimum in the SF mode dispersion (the spectra
at $\nu=2/5$ has been scaled down by a factor 2).The inset
compares polarized (P) and depolarized (D) spectra . b) Dispersion
of SF modes at $\nu = 2/5$ calculated by Mandal \emph{et
al.}\cite{mandal}. The inset shows the scheme of spin-split Landau
levels at $\nu$= 2/5. Landau levels are labelled  by quantum
numbers and arrows that indicate orientation of spin. Composite
fermions are shown as circles with two arrows that represent the
two vortices attached to each electron. }
\end{figure}
\par
Figure \ref{fig1/3}(a) displays spectra obtained at $\nu \gtrsim
1/3$ showing the long wavelength SW mode at E$_Z$ and a narrow
peak labelled SF$^+$ that occurs slightly below. The ordering of
lowest CF energy levels in the inset is consistent with Knight
shift results showing there is no change in spin polarization in
the range of our experiments \cite{khandelwal}. The SF$^+$ mode is
absent at $\nu = 1/3$ and grows as the filling factor increases.
This dependence suggests the mode is related to population of the
second CF Landau level $1 \uparrow$ as shown in the inset to the
figure. On this basis we assign the SF$^+$ peak to the SF
excitation due to $1 \uparrow \rightarrow 0 \downarrow$
transitions.
\par
\begin{figure}
\includegraphics{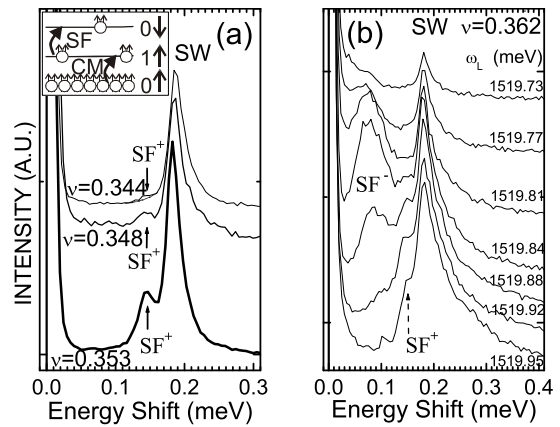}
\caption{\label{fig1/3} (a) Inelastic light scattering spectra at
filling factors $\nu \gtrsim 1/3$. Light polarizations are as in
Fig.\ref{fig2/5}. Vertical arrows show the peak assigned to the
spin-flip mode. The inset shows the three lowest CF levels near
1/3 and the transitions in SF modes and in charge density modes
(CM). (b) Resonant inelastic light scattering spectra at
$\nu$=0.362. Incident photon energies are given in meV. These
spectra display the coexistence of SF$^+$ and SF$^-$ modes that
occurs roughly in the middle of the filling factor range.}
\end{figure}
\par
The spectra in Figs. \ref{fig2/5} (a) and \ref{fig1/3} (a) confirm
the presence of the SF excitations predicted by the electron-like
level schemes shown in the insets. These results offer strong
evidence that the lowest CF levels are similar to those of
electrons. It is significant that spectra of SF excitations can be
measured over the whole filling factor range and that there is
markedly different filling factor dependence of SF modes near 2/5
and 1/3.
\par
\begin{figure}
\includegraphics{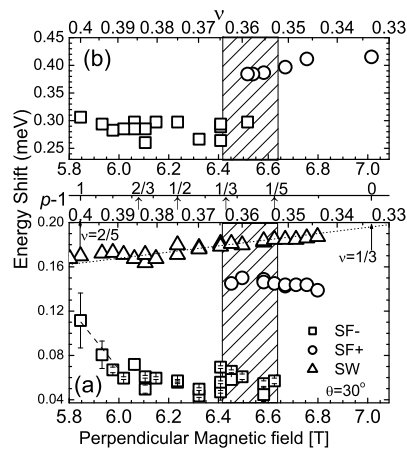}
\caption{\label{fig-energy} Energies of low-lying modes as
function of perpendicular field B. (a) Spin modes. The dashed line
is the E$_{SF}$ given by the variation of the CF cyclotron
frequency as discussed in the text. The dotted line is a fit of
the SW energy with E$_Z$ as described in the text. $p-1$ is the
population of the second LL. (b) Rotons of charge density modes.
The shaded area corresponds to the region where distinct modes
originating from 2/5 and 1/3 coexist.}
\end{figure}

Figure \ref{fig-energy}(a) plots the energies of the low-lying
spin modes. The SF peaks are quite different at the two filling
factor limits, with a marked change in character roughly in the
middle of the range. Near 2/5 the observed SF$^-$ mode is
interpreted as the roton minimum in the wave vector dispersion.
The position of the minimum is obtained by fitting the spectral
intensity below E$_Z$ with a roton density of states represented
as a convolution of a step function with a lorentzian function.
The mode energy collapses rapidly from 0.11meV until it stabilizes
at a very low value of 0.055meV. This mode disappears for fields
$\nu \lesssim 0.353$. Near 1/3 the very sharp SF$^+$ mode at
energy close to E$_Z$ has no dependence on magnetic field and
disappears for fields $\nu \gtrsim 0.363$.
\par
Figure \ref{fig-energy}(b) shows the magnetic field dependence of
the energies of the lowest charge-density excitations. These modes
arise from transitions $0 \uparrow \rightarrow 1 \uparrow$
labelled CM in the inset to Fig. \ref{fig1/3} (a), in which there
is no change in the spin degree of freedom. Near 1/3 the
excitation shown is the roton of the wave vector dispersion, and
near 2/5 it is the lowest of the two rotons \cite{moonsoo01}. The
large wave vector charge density modes, at E($\infty$), also seen
in the spectra, show similar behavior. The abrupt changes in the
energies of spin-flip and charge-density modes are in stark
contrast with the field dependence of the SW peak that is exactly
defined by the Zeeman energy E$_Z$ over the whole range.
\par
\par
The soft SF$^-$ mode collapses to a low energy that is
significantly smaller than E$_Z$ within a small magnetic field
interval $\Delta B < 0.2$T. In this narrow range of filling factor
the change in mode energy could be approximated by $\Delta
\omega_{c} = e \Delta B\ /c m_{CF}$, which ignores changes in
interactions among CF's. $\Delta \omega_{c}$ is obtained with a
determination of $m_{CF}$ from the energy E$_G(\infty)$ which
represents quasiparticle-quasihole pairs at large separation.
E$_G$ ($\infty$) is interpreted as the CF cyclotron energy given
by Eq.\ref{wc}. The result E$_G (\infty) = \hbar\omega_{c}$ =
0.64meV, reported at $\nu = 1/3$ by Kang \emph{et al.}
\cite{moonsoo01} for the same sample, yields a CF mass of $m_{CF}$
= 0.4$m_{o}$, where $m_{o}$ is the electron rest mass. The dashed
line in Fig.\ref{fig-energy} (a), which is a good representation
of the softening, is obtained with $m_{CF}$ = 0.4$m_{o}$ in
$\hbar\Delta \omega_{c}$.
\par
In spin-polarized states the spin reversal energy
E$^{\uparrow\downarrow}$ of CF  quasiparticles is a measure of
residual interactions. To obtain it from our measurements we
consider the SF$^+$ energy in the limit $\nu \rightarrow 1/3$ when
there is vanishing population of the  $1 \uparrow$ level ($p
\rightarrow 1$). In this limit coupling between the excited
quasiparticle and its quasihole is negligible and the SF
transition energy approaches the value:
\begin{equation}
\label{esf}
\rm{E_{SF}}\rightarrow\rm{E}_Z+\rm{E}^{\uparrow\downarrow}-\hbar\omega_{CF}
\end{equation}
Equation \ref{esf} and the determinations of E$_{SF}$ = 0.14 meV,
$\hbar\omega_{CF}$ = E$_G$ ($\infty$) = 0.64meV and E$_Z$ =
0.185meV yield E$^{\uparrow\downarrow}$ = 0.595meV. This
determination of E$^{\uparrow\downarrow}$ is consistent with
recent CF evaluations \cite{mandal,mandalactivation}.
\par
The direct access to low-lying collective modes between major states
of the FQHE offers venues to explore new behaviors of the liquids.
These states with partially populated CF levels may experience
further condensation under the impact of CF interactions, as
suggested by recent experimental and theoretical results
\cite{pan,izab,scar02}. It is intriguing that the SF$^-$ mode reaches
its lowest energy for filling factors $\nu \lesssim $ 0.385 where, as
indicated in Fig. \ref{fig-energy} (a), the filling of the $1
\uparrow$ level is $p-1= 2/3$. Here, the departure of the SF$^-$
energy from the prediction of Eq. (1) suggests an onset of new
interaction effects in the partially populated $1 \uparrow$ level.
The independence of the SF$^+$ energy on magnetic field that extends
to filling factors very close to 1/3 could also manifest the impact
of residual CF interactions when there is there is low population of
the $1\uparrow$ CF Landau level.
\par
Figure \ref{fig-energy} highlights the abrupt changes in the
energies of spin-flip and charge density modes that occur roughly
in the middle of the filling factor range. These results seem to
indicate the existence of two types of excitations arising either
from 1/3 or from 2/5 states. However, such picture can only be
regarded as incomplete at best. Figure \ref{fig1/3} (b) shows
spectra of SF modes when SF$^{+}$ and SF$^{-}$ modes coexist.
These spectra have a strong dependence on temperature. The
intensity of the higher energy mode SF$^{+}$ is seen to increase
markedly with simultaneous decrease in the intensity of the SF$^-$
mode as the temperature is increased above 200mK. Observations of
mode coexistence and its temperature dependence offer venues to
further studies of the impact of CF interactions.
\par
To conclude, resonant inelastic light scattering experiments
access low-lying excitations of electron liquids in the full range
of filling factors between states of the main sequence of the
FQHE. Near 2/5 and 1/3 we found evidence of lowest spin-split CF
Landau levels of charged fermions with spin 1/2, that are similar
to those of electrons. Spin-reversal energies determined from the
spectra reveal sizable composite fermion interactions. Current
evaluations of spin-reversed quasiparticle excitations are in
excellent quantitative agreement with our results. Interpretations
of low-lying excitation modes at filling factors in the middle of
the range between 1/3 and 2/5 require further experimental and
theoretical exploration.

\begin{acknowledgments}
We are grateful to J. K. Jain, S. H. Simon, H. L. Stormer and C. F.
Hirjibehedin for critical reading of the manuscript. This work was
supported in part by the Nanoscale Science and Engineering Initiative
of the National Science Foundation under NSF Award Number
CHE-0117752; and by a research grant of the W. M. Keck Foundation.
\end{acknowledgments}

\end{document}